\documentclass{article}
\usepackage{spconf,amsmath,amssymb,graphicx}
\usepackage{color}
\usepackage{balance}
\usepackage{parskip}

\usepackage[]{algorithm2e}
\usepackage{wasysym}
\usepackage{bm}
\usepackage{balance}
\usepackage{enumitem} % tune margin for \begin{itemize}
\usepackage{pifont}% http://ctan.org/pkg/pifont
\usepackage{multicol}

\def\thline{\noalign{\hrule height 1.0pt}}

\title{Class-conditional embeddings for music source separation}
\name{Prem Seetharaman$^{1,2}$, Gordon Wichern$^1$, Shrikant Venkataramani$^{1,3}$, Jonathan Le Roux$^1$\thanks{This work was performed while P. Seetharaman and S. Venkataramani were interns at MERL.}}
\address{$^1$Mitsubishi Electric Research Laboratories (MERL), Cambridge, MA, USA\\
$^2$Northwestern University, Evanston, IL, USA\\
$^3$University of Illinois at Urbana-Champaign, Champaign, IL, USA}

\begin{document}
\ninept
\maketitle
\begin{abstract}
Isolating individual instruments in a musical mixture has a  myriad of potential applications, and seems imminently achievable given the levels of performance reached by recent deep learning methods.  While most musical source separation techniques learn an independent model for each instrument, we propose using a common embedding space for the time-frequency bins of all instruments in a mixture inspired by deep clustering and deep attractor networks.  Additionally, an auxiliary network is used to generate parameters of a Gaussian mixture model (GMM) where the posterior distribution over GMM components in the embedding space can be used to create a mask that separates individual sources from a mixture.  In addition to outperforming a mask-inference baseline on the MUSDB-18 dataset, our embedding space is easily interpretable and can be used for query-based separation.
\end{abstract}
\section{Introduction}
Audio source separation is the act of isolating sound-producing sources in an auditory scene. Examples include separating singing voice from accompanying music, the voice of a single speaker at a crowded party, or the sound of a car backfiring in a loud urban soundscape. Recent deep learning techniques have rapidly advanced the performance of such source separation algorithms leading to state of the art performance in the separation of music mixtures~\cite{Takahashi2018}, separation of speech from non-stationary background noise~\cite{Erdogan2015}, and separation of the voices from simultaneous overlapping speakers~\cite{Hershey2016}, often using only a single audio channel as input, i.e., no spatial information.  

In this work we are concerned with separation networks that take as input a time-frequency (T-F) representation of a signal (e.g., magnitude spectrogram), and either predict the separated source value in each T-F bin directly or via a T-F mask that when multiplied with the input recovers the separated signal. An inverse transform is then used to obtain the separated audio.  One approach for training such algorithms uses some type of signal reconstruction error, such as the mean square error between magnitude spectra \cite{Erdogan2015, Wang2017Overview}.  An alternative approach referred to as deep clustering \cite{Hershey2016, isik2016single, Wang2018ICASSP} uses affinity-based training by estimating a high-dimensional embedding for each T-F bin, training the network with a loss function such that the embeddings for T-F bins dominated by the same source should be close to each other and those for bins dominated by different sources should be far apart.  This affinity-based training is especially valuable in tasks such as speech separation, as it avoids the permutation problem during network training where there is no straightforward mapping between the order of targets and outputs.

\begin{figure}
    \centering
    \includegraphics[width=\linewidth]{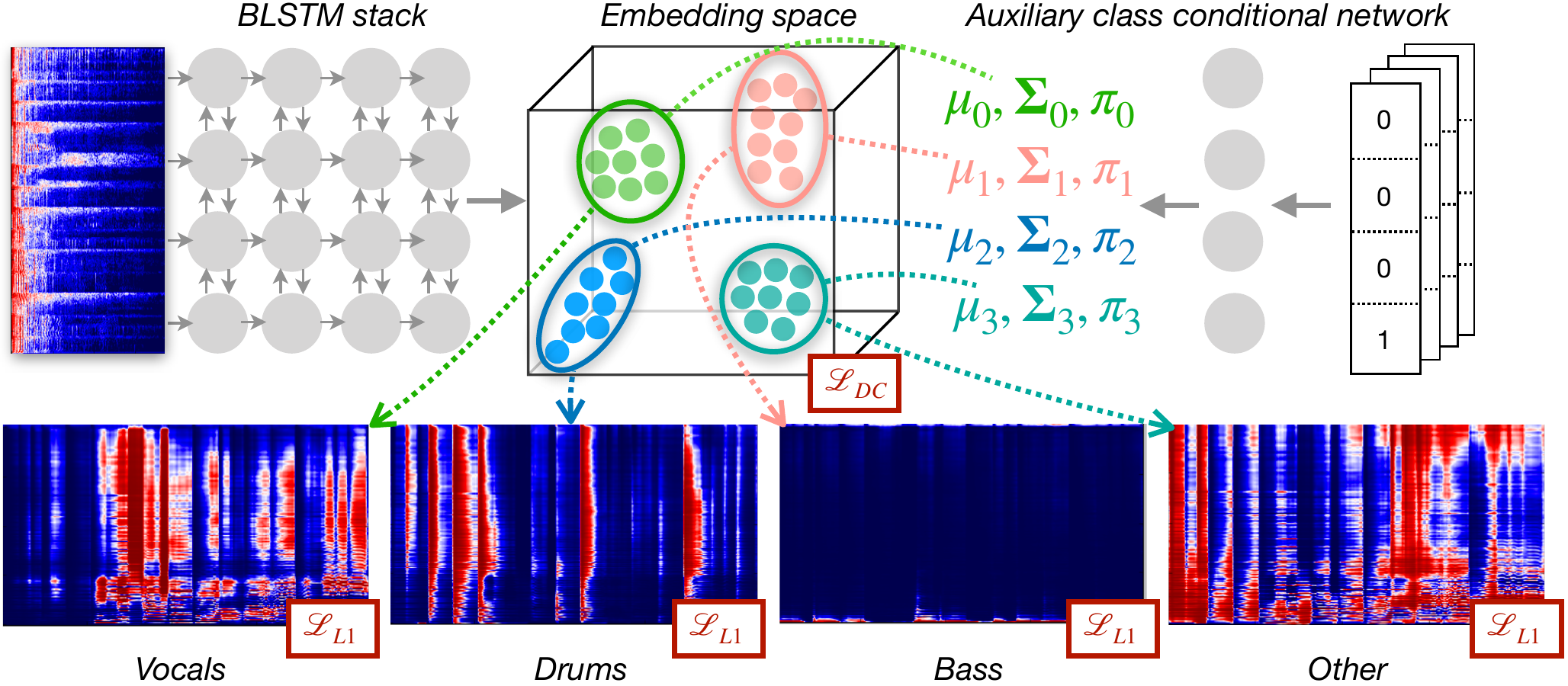}
    \caption{Class-conditional embedding network architecture. An auxiliary network generates the parameters of Gaussians in the embedding space, taking a one-hot vector indicating the class as input. The masks are generated using the posteriors of the Gaussian mixture model across all the sources. The network is trained using a deep clustering loss ($\mathcal{L}_{DC}$) on the embedding space and an $L^1$ loss ($\mathcal{L}_{L^1}$) on the masked spectrograms.  }
    \label{fig:block_diagram}
\end{figure}

Deep clustering for music source separation was previously investigated with Chimera networks for singing voice separation in~\cite{luo2017deep}. Chimera networks \cite{luo2017deep, Wang2018ICASSP} have multiple parallel output heads which are trained simultaneously on different tasks.  Specifically, in the case of singing voice separation, one output head is used to directly approximate the soft mask for extracting vocals, while the other head outputs an embedding space that optimizes the deep clustering loss. When both heads are trained together, results are better than using any single head alone.  Another approach for combining deep clustering and mask-based techniques was presented in \cite{isik2016single} where a deep clustering network, unfolded k-means layers, and a second stage enhancement network are trained end-to-end.  The deep attractor network~\cite{luo2018speaker} computes an embedding for each T-F bin similar to deep clustering, but creates a mask based on the distance of each T-F bin to an attractor point for each source.  The attractors can either be estimated via k-means clustering, or learned as fixed points during training.

In this work, we consider deep attractor-like networks for separating multiple instruments in music mixtures.  While embedding networks have typically been used in speech separation where all sources in a mixture belong to the same class (human speakers), we extend the formulation to situations where sources in a mixture correspond to distinct classes (e.g., musical instruments). Specifically, our class-conditional embeddings work as follows: first, we propose using an auxiliary network to estimate a Gaussian distribution (mean vector and covariance matrix) in an embedding space for each instrument class we are trying to separate.   Then, another network computes an embedding for each T-F bin in a mixture, akin to deep clustering. Finally, a mask is generated based on the posterior distribution over classes for each T-F bin. The network can be trained using a signal reconstruction objective, or in a multi-task (Chimera-like) fashion with an affinity-based deep clustering loss used as a regularizer.  

Deep clustering and deep attractor networks typically focus on speaker-independent speech separation where the mapping of input speaker to output index is treated as a nuisance parameter handled via permutation free training \cite{Hershey2016, isik2016single, Kolbæk2017}.  Several recent works on speaker-conditioned separation \cite{delcroix2018single, wang2018deep} allow separation of a targeted speaker from a mixture in a manner similar to how we extract specific instruments.  Learning an embedding space for speaker separation that could also be used for classification was explored in recent work \cite{drude2018deep}. However, their work did this by introducing a classification-based loss function. Here, the conditioning is introduced as input into the network rather than output from the network. Regarding specific musical instrument extraction from mixtures, a majority of methods \cite{huang2014singing, nugraha2016multichannel, uhlich2017improving, Takahashi2018} use an independent deep network model for each instrument, and then combine these instrument specific network outputs in post-processing using a technique such as the multi-channel Wiener filter \cite{nugraha2016multichannel, uhlich2017improving}.  While the efficacy of independent instrument modeling for musical source separation was confirmed by the results of a recent challenge \cite{stoter2018sisec}, the requirements both in terms of computational resources and training data can be large, and scaling up the number of possible instruments can be prohibitive. 

Recently, the work in \cite{kumar2018music} demonstrated that a common embedding space for musical instrument separation using various deep attractor networks could achieve competitive performance.  Our system is similar to the anchored and/or expectation-maximization deep attractor networks in \cite{kumar2018music}, but we use an auxiliary network to estimate the mean and covariance parameters for each instrument.  We also explore what type of covariance model is most effective for musical source separation (tied vs.\ untied across classes, diagonal vs.\ spherical). Furthermore, we discuss a simple modification of our pre-trained embedding networks for query-by-example separation \cite{pardo2006finding, el2017fly, ozerov2017comparative}, where given an isolated example of a sound we want to separate, we can extract the portion of a mixture most like the query without supervision.

\section{Embedding Networks}

Let $\mathbf{X} \in \mathbb{C}^{F\times T}$ be the complex spectrogram of the mixture of $C$ sources $\mathbf{S_c} \in \mathbb{C}^{F\times T}$ for $c=1,\dots,C$.  An embedding network computes
\begin{equation}
    \mathbf{V}=f(\tilde{\mathbf{X}})
\end{equation}
where $\tilde{\mathbf{X}}$ is the input feature representation (we use the log-magnitude spectrogram in this work), and $\mathbf{V} \in \mathbb{R}^{FT\times K}$ contains a $K$-dimensional embedding for each T-F bin in the spectrogram.  The function $f$ is typically a deep neural network composed of bidirectional long short-term memory (BLSTM) layers, followed by a dense layer.  We then create a mask for each source $\mathbf{m_c} \in \mathbb{R}^{FT\times 1}$, with
\begin{equation}
\sum_{c=1}^C m_{c,j}=1,\quad j=1,\dots,FT
\label{eq:softmask}
\end{equation}
from $\mathbf{V}$.  Deep clustering \cite{Hershey2016, isik2016single} builds binary masks via k-means clustering on $\mathbf{V}$ (soft masks can also be obtained via soft k-means), and is trained by minimizing the difference between the true and estimated affinity matrices,  
\begin{equation}
\mathcal{L}_{\text{DC}}(\mathbf{V},\mathbf{Y}) = \| \mathbf{V}\mathbf{V}^T - \mathbf{Y}\mathbf{Y}^T\|_{F}^2
\label{eq:dpcl_loss}
\end{equation}
where $\mathbf{Y}\in \mathbb{R}^{TF\times C}$ indicates which of the $C$ sources dominates each T-F bin.  Deep attractor networks \cite{luo2018speaker} use the distance between the embeddings and fixed attractor points in the embedding space to compute soft masks, and are typically trained with a signal reconstruction loss function, such as the $L^1$ loss between the estimated and ground truth magnitude spectrograms
\begin{equation}
\mathcal{L}_{L^1} = \sum_{c=1}^C|\mathbf{m}_c \odot \mathbf{x} - \mathbf{s}_c|.
\label{eq:mi_loss}
\end{equation}
where $\mathbf{x},\mathbf{s}_c\in \mathbb{R}^{FT\times 1}$ are the flattened spectrogram of the mixture, and ground truth source, respectively.  We can obtain the separated time domain signal from the estimated magnitude $\hat{\mathbf{s}}_c=\mathbf{m}_c \odot \mathbf{x}$ after an inverse STFT using the mixture phase.

Chimera networks \cite{luo2017deep, Wang2018ICASSP} combine signal reconstruction and deep clustering losses, using two heads stemming from the same underlying network (stacked BLSTM layers). In this work, we also combine the loss functions from \eqref{eq:dpcl_loss} and \eqref{eq:mi_loss} (with equal weighting), but the gradients from both propagate into the same embedding space, rather than separate heads.

\section{Conditioning embeddings on class}
When we are interested in separating sources that belong to distinctly different groups, i.e., classes, each source $c$ has an associated class label $z_c$, and we assume here that each mixture contains at most one isolated source per class label.  Estimating the mask $m_{c,j}$ in (\ref{eq:softmask}) for source (class) $c$ and T-F bin $j$ is then equivalent to estimating the posterior over classes $p(z_c|\mathbf{v}_j)$ given the corresponding $K$-dimensional network embedding.  For simplicity we use a Gaussian model of the embedding space
and obtain the mask from
\begin{equation}
    m_{c,j}=p(z_c|\mathbf{v}_j)=\frac{\pi_c\ \mathcal{N}(\mathbf{v}_j|\pmb{\mu}_c, \mathbf{\Sigma_c})}{\sum_{i=1}^{C}\pi_i\ \mathcal{N}(\mathbf{v}_j|\pmb{\mu}_i, \mathbf{\Sigma_i})}.
    \label{eq:postmask}
\end{equation}
The Gaussian parameters $(\pmb{\mu}_c, \mathbf{\Sigma_c})$ and class prior $\pi_c$ for each class are learned end-to-end along with the embedding network. The generation of the parameters of each Gaussian from the auxiliary class-conditional network is the maximization step in the expectation-maximization (EM) algorithm (trained through stochastic gradient descent), and the generation of the mask is the expectation step. Rather than unfolding a clustering algorithm as in \cite{isik2016single}, we instead can learn the parameters of the clustering algorithm efficiently via gradient descent. Further, the soft mask is generated directly from the posteriors of the Gaussians, rather than through a second-stage enhancement network as in \cite{isik2016single}. A diagram of our system can be seen in Fig.~\ref{fig:block_diagram}.

We also draw a connection between the class-conditional masks of \eqref{eq:postmask} and the adaptive pooling layers for sound event detection in \cite{mcfee2018adaptive}, which are also conditioned on class label. In \cite{mcfee2018adaptive}, an activation function that is a variant of softmax
with learnable parameter $\alpha$ is introduced. If $\alpha$ is very high, the function reduces to a max function, heavily emphasizing the most likely class. If it is low, energy is spread more evenly across classes approximating an average.
Our work uses a similar idea for source separation. A softmax nonlinearity is comparable to the posterior probability computation used in the expectation step of the EM algorithm in our Gaussian mixture model (GMM). For a GMM with tied spherical covariance, $\alpha$ 
is the inverse of the variance. A similar formulation of softmax was also used in \cite{isik2016single}, where k-means was unfolded on an embedding space. In that work $\alpha$ was set manually to a high value for good results. In our work, we effectively learn the optimal $\alpha$ (the inverse of the covariance matrix) for signal reconstruction rather than setting it manually, but still conditioning it on class as in \cite{mcfee2018adaptive} for source separation.

\section{Experiments}
Our experiment is designed to investigate whether our proposed class-conditional model outperforms a baseline mask inference model. We also explore which covariance type is most suitable for music source separation. We do this by evaluating the SDR of separated estimates of vocals, drums, bass, and other in the MUSDB \cite{musdb18} corpus using the museval package\footnote{https://github.com/sigsep/sigsep-mus-eval}. Finally, we show the potential of our system to perform querying tasks with isolated sources. 

\begin{figure}
    \centering
    \includegraphics[width=\linewidth]{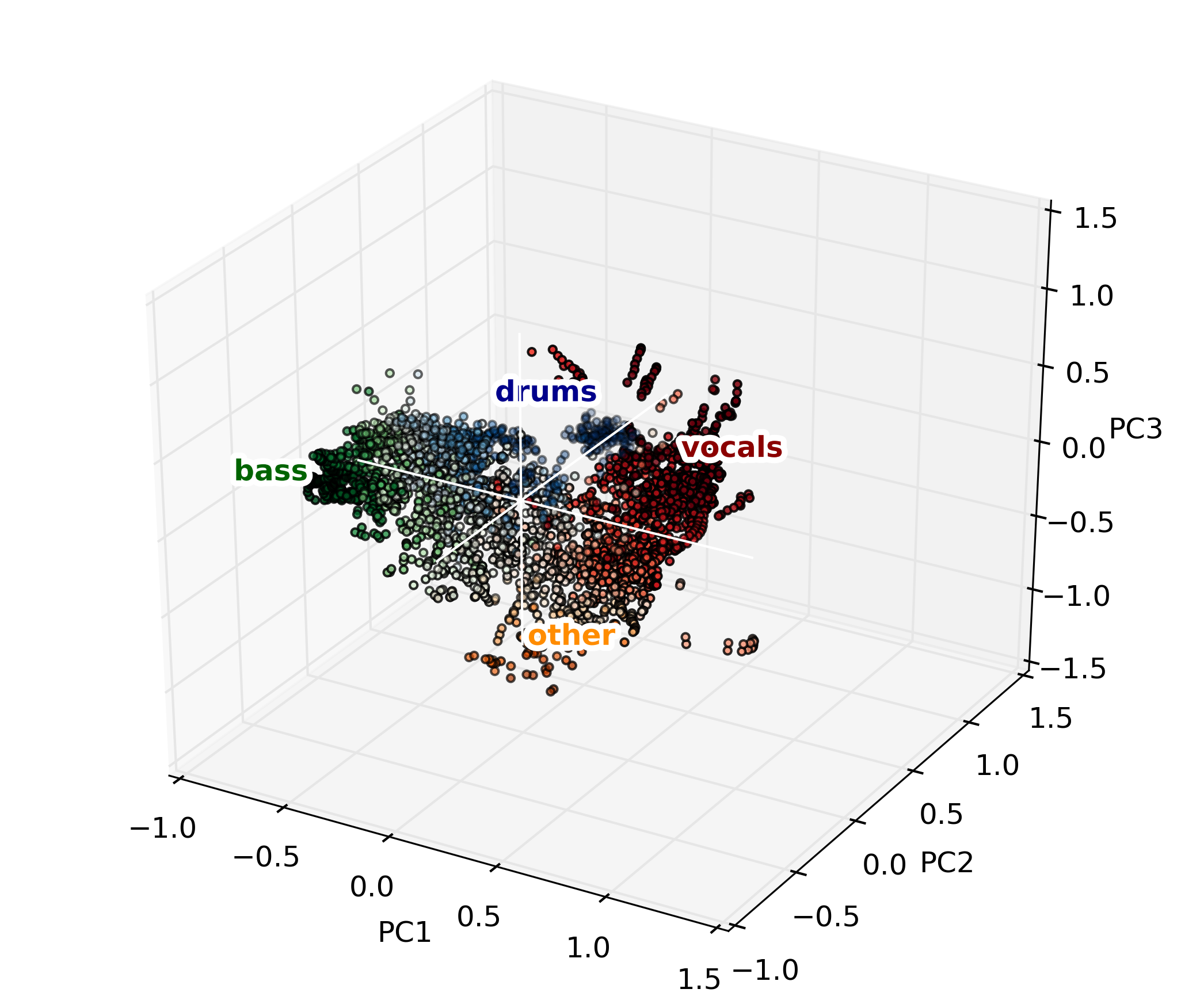}
    \caption{Visualization of the embedding space learned for music source separation with the tied spherical covariance model.}
    \label{fig:embedding}
\end{figure}

\subsection{Dataset and training procedure}
We extend Scaper \cite{salamon2017scaper}, a library for soundscape mixing designed for sound event detection, to create large synthetic datasets for source separation. We apply our variant of Scaper to the MUSDB training data, which consists of 100 songs with vocals, drums, bass, and other stems to create $20000$ training mixtures and $2000$ validation mixtures, all of length $3.2$ seconds at a sampling rate of $48$ kHz. Of the $100$ songs, we use $86$ for training and $14$ for validation. The remaining $50$ songs in the MUSDB testing set are used for testing. The training and validation set mixtures are musically incoherent (randomly created using stems from different songs) and each contains a random $3.2$ second excerpt from a stem audio file in MUSDB (vocals, drums, bass, and other). All four sources are present in every training and validation mixture. 

Time domain stereo audio is summed to mono and transformed to a single-channel log-magnitude spectrogram with a window size of $2048$ samples ($\approx 43$ ms) and a hop size of $512$ samples. Our network consists of a stack of 4 BLSTMs layers with $300$ units in each direction for a total of $600$. Before the BLSTM stack, we project the log-magnitude spectrogram to a mel-spectrogram with $300$ mel bins. The mel spectrogram frames are fed to the BLSTM stack which projects every time-mel bin to an embedding with $15$ dimensions. 

The auxiliary class-conditional network takes as input a one-hot vector of size $4$, one for each musical source class in our dataset. It maps the one-hot vector to the parameters of a Gaussian in the embedding space. For an embedding space of size $K$, and diagonal covariance matrix, the one-hot vector is mapped to a vector of size $2K+1$: $K$ for the mean, $K$ for the variance, and $1$ for the prior. After the parameters of all Gaussians are generated, we compute the mask from the posteriors across the GMM using Eq.~\eqref{eq:postmask}. The resultant mask is then put through an inverse mel transform to project it back to the linear frequency domain, clamped between $0$ and $1$, and applied to the mixture spectrogram. The system is trained end to end with $L^1$ loss and the embedding space is regularized using the deep clustering loss function. To compute the deep clustering loss, we need the affinity matrix for the mel-spectrogram space. This is computed by projecting the ideal binary masks for each source into mel space and clamping between $0$ and $1$. The deep clustering loss is only applied on bins that have a log magnitude louder than $-40$ db, following \cite{Hershey2016}.

We evaluate the performance of multiple variations of class-conditional embedding networks on the MUSDB18 \cite{musdb18} dataset using source-to-distortion ratio (SDR)\footnote{https://github.com/sigsep/sigsep-mus-eval}. At test time, we apply our network to both stereo channels independently and mask the two channels of complex stereo spectrogram. We explore several variants of our system, specifically focusing on the possible covariance shapes of the learned Gaussians. We compare these models to a baseline model that is simply a mask inference network (the same BLSTM stack) with $4F$ outputs (one mask per class) followed by a sigmoid activation. All networks start from the same initialization and are trained on the same data.

\subsection{Results}
Table~\ref{tab:results} shows SDR results for the baseline model and the four covariance model variants. We find that all four of our models that use an embedding space improve significantly on the baseline for vocals and other sources. The best performing model is a GMM with tied spherical covariance, which reduces to soft k-means, as used in \cite{isik2016single}. The difference here is that the value for the covariance is learned rather than set manually. The covariance learned was $0.16$, or an $\alpha$ value of $6$, close to the $\alpha$ value of $5$ found in \cite{isik2016single}. The embedding space for this model on a sample mixture is visualized in Fig.~\ref{fig:embedding} using Principal Component Analysis. We observe that there exist ``bridges'' between some of the sources. For example, other and vocals share many time-frequency points, possibly due to their source similarity. Both sources contain harmonic sounds and sometimes leading melodic instruments. However, unlike other embedding spaces (e.g., word2vec) where things that are similar are near each other in the embedding space, we instead have learned a separation space, where sources that are similar (but different) seem to be placed far from each other in the embedding space. We hypothesize that this is to optimize the separation objective. In \cite{luo2018speaker}, it is observed that attractors for speaker separation come in two pairs, across from each other. Our work suggests that the two pairs may correspond to similar sources (e.g., separating female speakers from one another and separating male speakers from one another). Verifying this and understanding embedding spaces learned by embedding networks will be the subject of future work.

\begin{figure}
    \centering
    \includegraphics[width=\linewidth]{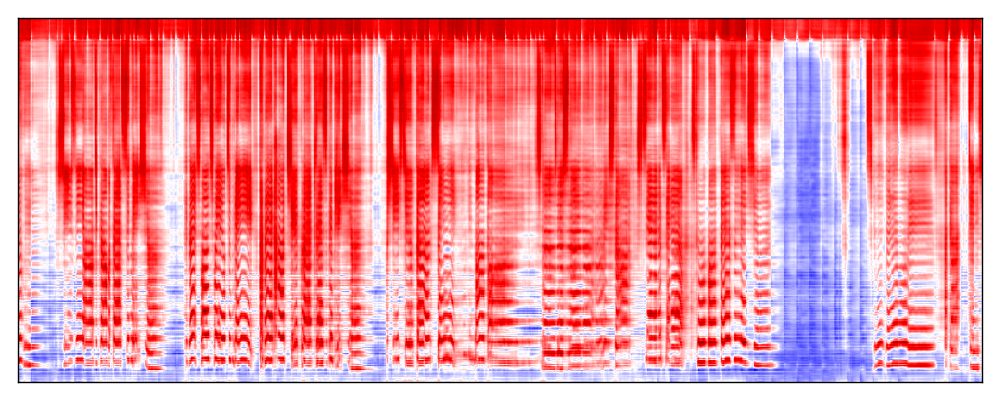}
    \caption{An example of an embedding dimension for a GMM with diagonal covariance. The embedding selected here is the one with the lowest variance ($0.04$) for the vocals source.}
    \label{fig:embedding_dim}
\end{figure}

 We hypothesize that the reason the simplest covariance model (tied spherical) performs best in Table~\ref{tab:results} is that for the diagonal case, the variances collapse in all but a few embedding dimensions. Embedding dimensions with the lowest variance contribute most to the overall mask.  As a result, they essentially become the mask by themselves, reducing the network more to mask inference rather than an embedding space. An example of this can be seen in Fig.~\ref{fig:embedding_dim}, where the embedding dimension has essentially reduced to a mask for the vocals source. With a spherical covariance model, each embedding dimension must be treated equally, and the embedding space cannot collapse to mask inference. A possible reason tied spherical performs better than untied spherical, may be that the network becomes overly confident (low variance) for certain classes. With a tied spherical covariance structure, all embedding dimensions and instrument classes are equally weighted, forcing the network to use more of the embedding space, perhaps leading to better performance.

\begin{table}[]
\centering
\caption{SDR comparing mask inference baseline (BLSTM) with deep clustering (DC) plus GMM models with different covariance types (diagonal, tied diagonal, spherical, and tied spherical).
}
\label{tab:results}
\begin{tabular}{l|cccc}
\thline
Approach                        & Vocals & Drums  & Bass   & Other  \\ \thline
BLSTM                           & $3.82$ & $4.14$ & $2.48$ & $2.35$ \\ \hline
DC/GMM - diag. (untied) & $4.20$ & $\mathbf{4.26}$ & $2.58$ & $2.55$ \\ %\hline
DC/GMM - diag. (tied)   & $4.04$ & $3.96$ & $2.48$ & $2.47$ \\ %\hline
DC/GMM - sphr. (untied) & $4.21$ & $4.19$ & $2.29$ & $\mathbf{2.58}$ \\ %\hline
DC/GMM - sphr. (tied)   & $\mathbf{4.49}$ & $4.23$ & $\mathbf{2.73}$ & $2.51$ \\ \thline
\end{tabular}\vspace{-.3cm}
\end{table}

\subsection{Querying with isolated sources}
To query a mixture with isolated sources, we propose a simple approach that leverages an already trained class-conditional embedding network. We take the query audio and pass it through the network to produce an embedding space. Then, we fit a Gaussian with a single component to the resultant embedding space. Next, we take a mixture that may contain audio of the same type as the query, but not the exact instance of the query, and pass that through the same network. This produces an embedding for the mixture. To extract similar content to the query from the mixture, we take the Gaussian that was fit to the query embeddings and run the expectation step of EM by calculating the likelihood of the mixture's embeddings under the query's Gaussian. Because there is only one component in this mixture model, calculating posteriors  gives a mask of all ones. To alleviate this, we use the likelihood under the query Gaussian as the mask on the mixture and normalize it to $[0,1]$ by dividing each likelihood by the maximum observed likelihood value in the mixture.

\begin{figure}
    \centering
    \includegraphics[width=\linewidth]{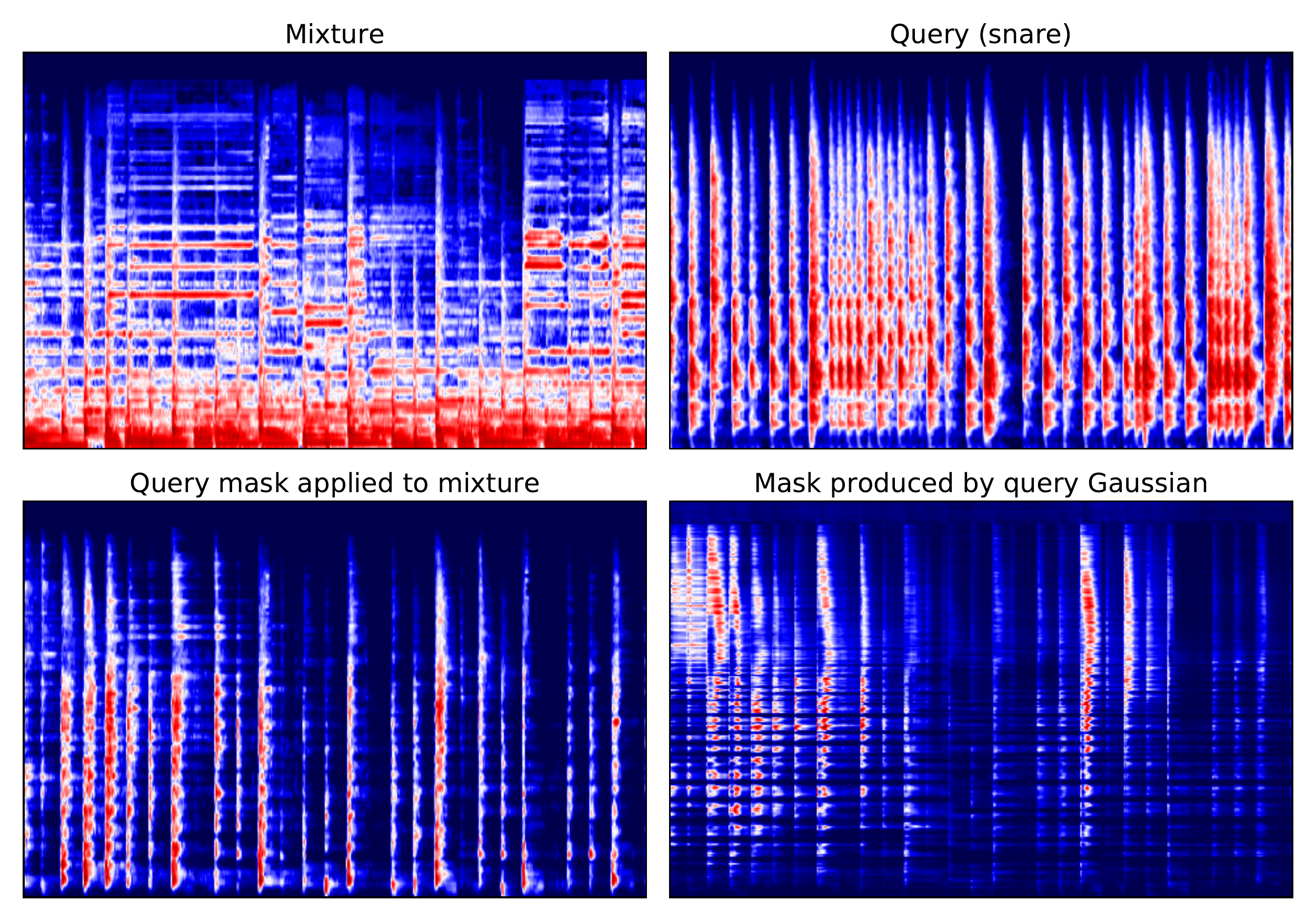}
    \caption{An example of query by isolated source using our trained tied-spherical model.  Upper left: mixture spectrogram, upper right: query (snare drum) spectrogram, bottom right: query mask, bottom left: masked mixture using query mask.}
    \label{fig:query}
\end{figure}

An example of query by isolated source can be seen in Fig.~\ref{fig:query}. We use a recording of solo snare drum as our query. The snare drum in the query is from an unrelated recording found on YouTube. The mixture recording is of a song with simultaneous vocals, drums, bass, and guitar (\textit{Heart of Gold} - Neil Young). The Gaussian is fit to the snare drum embeddings and transferred to the mixture embeddings. The mask produced is similar to the query as is the extracted part of the mixture. This invariance of embedding location was a result of conditioning the embeddings on the class. 

\section{Conclusion}
We have presented a method for conditioning on class an embedding space for source separation. We have extended the formulation of deep attractor networks and other embedding networks to accommodate Gaussian mixture models with different covariances. We test our method on musical mixtures and found that it outperforms a mask inference baseline. We find that the embeddings found by the network are interpretable to an extent and hypothesize that embeddings are learned such that source classes that have similar characteristics are kept far from each other in order to optimize the separation objective. Our model can be easily adapted to a querying task using an isolated source. In future work, we hope to investigate the dynamics of embedding spaces for source separation, apply our approach to more general audio classes, and explore the querying task further.

\bibliographystyle{IEEEtran_nourl}
\bibliography{refs}

\end{document}